\documentstyle{mn} 

\input epsf

\def\plotone#1{\centering \leavevmode
\epsfxsize=\columnwidth \epsfbox{#1}}

\begin{document}

\title[Polarization in galaxy clusters induced by the CMB quadrupole]
{Microwave polarization in the direction of galaxy clusters induced by
the CMB quadrupole anisotropy}

\author[S. Y. Sazonov and R. A. Sunyaev]{S. Y.~Sazonov$^{1,2}$ and
R. A.~Sunyaev$^{1,2}$\\
$^1$MPI f\"ur Astrophysik, Karl-Schwarzschild-Str.~1, 86740 Garching bei
M\"unchen, Germany\\
$^2$Space Research Institute (IKI), Profsouznaya~84/32, Moscow 117810,
Russia}
\maketitle

\begin{abstract}
Electron scattering induces a polarization in the cosmic
microwave background (CMB) signal measured in the direction of a
galaxy cluster due to the presence of a quadrupole component in the
CMB temperature distribution. Measuring the polarization towards
distant clusters provides the unique opportunity to observe the
evolution of the CMB quadrupole at moderate redshifts,
$z\sim$~0.5--3. We demonstrate that for the local cluster population
the polarization degree will depend on the cluster celestial
position. There are two extended regions in the sky, which are opposite to
each other, where the polarization is maximal,
$\sim0.1(\tau/0.02)$~$\mu$K in the Rayleigh-Jeans part of the CMB
spectrum ($\tau$ being the Thomson optical depth across the cluster)
exceeding the contribution from the cluster transverse peculiar motion if
$v_t<1300$~km/s. One can hope to detect this small signal by measuring
a large number of clusters, thereby effectively removing the systematic
contribution from other polarization components produced in
clusters. These polarization effects, which are of the order of
$(v_t/c)^2\tau$, $(v_t/c)\tau^2$ and $(kT_e/m_ec^2)\tau^2$, as well as
the polarization due to the CMB quadrupole, were previously calculated
by Sunyaev and Zel'dovich for the Rayleigh-Jeans region. We fully
confirm their earlier results and present exact frequency dependencies
for all these effects. The polarization is considerably higher in the
Wien region of the CMB spectrum.
\end{abstract}

\section{INTRODUCTION}
The intensity of the cosmic microwave background radiation is
distorted in the direction of galaxy clusters. There are two basic
effects of this kind: the thermal effect \cite{sunyaev_zeldovich72} 
and the kinematic effect \cite{sunyaev_zeldovich80}. The
polarization of the CMB should be changed due to the presence of
galaxy clusters in the line of sight as well
\cite{sunyaev_zeldovich80,zeldovich_sunyaev80,sunyaev_zeldovich81}. In fact
there are several separate polarization effects but since 
the expected amplitude of each of them is very small, until recently
there was little interest to them. However, we have observed such a
significant progress in CMB measurements over the period passed since
1980, that the possibilty to detect a CMB-induced polarization signal
from galaxy clusters seems no longer unrealistic. Furthermore, the CMB
experiments planned for the next decade, which include the space
missions MAP (http://map.gsfc.nasa.gov), SPOrt on the International
Space Station \cite{cortiglioni99}, and Planck Surveyor
(http://astro.estec.esa.nl/Planck; De Zotti et al. 1999), and a number
of ground-based and baloon-borne experiments, e.g. POLAR
\cite{keating98}, will already reach a level of sensitivity which is
not far from what is needed for such a detection. This drastic change
in the situation is reflected in the fact that several new papers
(Audit \& Simmons 1998; Itoh, Nozawa \& Kohyama 1998; Hansen \& Lilje
1999) discussing the polarization effects in clusters have appeared
over the past year or so. Some authors \cite{audit_simmons98,itoh98}
have claimed the necessity to revise the original results of Sunyaev
\& Zel'dovich \shortcite{sunyaev_zeldovich80}. This gave us a first
motivation to return to this problem.
 
Another reason for us to begin this study was the discovery by COBE
(see, e.g. Kogut et al. 1996) of a quadrupole component in the CMB
angular anisotropy. Zel'dovich \& Sunyaev
\shortcite{zeldovich_sunyaev80} showed that the presence of a
quadrupole component should cause a change in the CMB
polarization in the direction of a galaxy cluster. Already our first
estimate based on the COBE measurement lead us to the rather surprising
conclusion that this polarization effect should generally be stronger than
any of the other polarization effects relating to galaxy
clusters. This fact makes the quadrupole-induced polarization
potentially very important for cosmology. Indeed, measuring the
microwave polarization signal towards distant galaxy clusters opens the
unique opportunity to effectively probe the CMB quadrupole anisotropies at
intermediate redshifts, $z\sim$~0.5--3, i.e. from positions in the
Universe different from ours, thus giving us a way to reduce the
cosmic-variance uncertainty \cite{kamionkowski_loeb97}. Furthermore,
by observing the effect in clusters (or any large-scale ionized
gas clouds) at large $z$ we could be able to follow the 
evolution of the CMB quadrupole, which should be different for
universes with different cosmological parameters.

The structure of the paper is as follows. In Section 2 we give the
basic formulae which allow one to find the angular distribution of
the polarized radiation resulted from scattering of a given initial angular
distribution of unpolarized radiation by a single resting electron. In Section
3 we calculate the amplitude and distribution over the sky of the
polarization imposed by the CMB quadrupole on the signal from clusters. We
study other polarization effects connected with galaxy clusters in Section 4,
which is followed by a Discussion of possible ways to detect the cluster
polarization due to the CMB quadrupole.

\section{POLARIZATION DUE TO SCATTERING OF ANISOTROPIC RADIATION BY A
SINGLE ELECTRON}

In this section, which lays the basis for our further analysis, we
basically repeat the original treatment of Zel'dovich \& Sunyaev
\shortcite{zeldovich_sunyaev80}, which is also described in detail in
a review paper of Sunyaev \& Zel'dovich \shortcite{sunyaev_zeldovich81}.

We consider the situation that a single electron being at rest scatters
initially unpolarized radiation. The frequency of a photon does not  
change due to scattering (one can safely neglect the small change due to recoil
when dealing with the CMB), and we therefore can, for the moment,
ignore the spectral dependence of the radiation and consider only its
angular distribution. This distribution can, in general, be presented
as the sum of components proportional to the Legendre polynomials
$P_n(\mu)$: $1$, $\mu$, $(\mu^2-1/3)$, etc., where $\mu$ is the
cosine of the angles between the wave vector of the incident photon
and one of a set of directions that uniquely define our particular angular
distribution. The incident radiation is described by a usual set of
Stokes parameters, ($I$, $Q$, $U$, $V$), where $Q=U=V=0$. Our aim is
to determine the corresponding quantities (to be marked with
primes) for the scattered radiation. The fourth Stokes parameter,
$V$, describes circular polarization which is not produced in a
scattering and will, therefore, not be mentioned further. We define
the Stokes parameters of the scattered radiation with respect to the
plane that contains a) the specific direction about which a given
radiation anisotropy component has an axial symmetry (see above), and
b) the wave vector of the scattered photon, so $Q>0$, $U=0$
corresponds to a linear polarization in the direction perpendicular to
this plane. With this choice of the reference frame, it is easy to
show \cite{chandrasekhar50,sunyaev_zeldovich81} that the
scattered radiation will be
 
\begin{equation}
I^\prime(\mu^\prime)=\int_{-1}^{1}\left[\frac{1}{2}+\frac{3}{16}(3\mu^{\prime
2}-1)\left(\mu^2-\frac{1}{3}\right)\right] I(\mu)\,d\mu,
\label{prob_i}
\end{equation}

\begin{equation}
Q^\prime(\mu^\prime)=\int_{-1}^{1}\frac{9}{16}(1-\mu^{\prime
2})\left(\mu^2-\frac{1}{3}\right) I(\mu)\,d\mu,\,\,\,U^\prime(\mu^\prime)=0.
\label{prob_qu}
\end{equation}
One can see that the angular scattering function for the $Q$
Stokes parameter is of strictly quadrupole form (with respect to
$\mu$). This fact is of importance for our further consideration,
because it means that, due to the orthogonality condition for Legendre
polynomials, only the quadrupole component, neither the dipole nor the
higher harmonics, in the incident intensity distribution will
contribute to $Q^\prime$.

Now let the isotropic and multipole components of the intensity
distribution be $I_0$, $I_1\mu$, $I_2(\mu^2-1/3)$, etc. Then upon
integration in (\ref{prob_i})--(\ref{prob_qu}) we derive

\begin{equation}
I^\prime(\mu^\prime)=I_0+\frac{1}{10}I_2\left(\mu^{\prime
2}-\frac{1}{3}\right),
\label{quad_i}
\end{equation}

\begin{equation}
Q^\prime(\mu^\prime)=\frac{1}{10}I_2(1-\mu^{\prime
2}),\,\,\,U^\prime(\mu^\prime)=0.
\label{quad_qu}
\end{equation}
As expected, the isotropic component remains unchanged upon
scattering. If one now considers a situation that the optical depth of
the scattering medium is very small, $\tau\ll 1$, then only single
scatterings of photons will play role, and it is straightforward to
modify formulae (\ref{quad_i}), (\ref{quad_qu}), which were obtained
for an individual electron, for this case:

\begin{eqnarray}
I^\prime(\mu^\prime)=I(\mu^\prime)(1-\tau)+\tau\left[I_0+\frac{1}{10}I_2\left(\mu^{\prime2}-\frac{1}{3}\right)\right]
\nonumber\\
=I(\mu^\prime)-\tau\left[I_1\mu^\prime+0.9
I_2\left(\mu^{\prime2}-\frac{1}{3}\right)+\sum_{n=3}^{\infty}I_n
P_n(\mu^\prime)\right],
\label{quad_i_tau}
\end{eqnarray} 

\begin{equation}
Q^\prime(\mu^\prime)=\frac{1}{10}\tau I_2(1-\mu^{\prime
2}),\,\,\,U^\prime(\mu^\prime)=0.
\label{quad_qu_tau}
\end{equation}
Equations (\ref{quad_i_tau}), (\ref{quad_qu_tau}) show that electron
scattering tends to smooth anisotropies in the intensity distribution
and to produce a polarization on the order of $Q^\prime/I^\prime=0.1\tau
I_2/I_0(1-\mu^{\prime 2})$ \cite{zeldovich_sunyaev80,sunyaev_zeldovich81}.

\section{POLARIZATION DUE TO THE CMB QUADRUPOLE}

The COBE satellite provided the first measurement of the quadrupole
component in the CMB angular distribution. Although the accuracy of
this measurement is very low -- its amplitude, $Q_{rms}$, is said
to lie in the range $[4,28]$~$\mu$K at 95\% confidence \cite{kogut96} --
it is very likely that the detection of the quadrupole is 
real, because the presence of a similar quadrupole signal,
$Q_{rms-ps}\sim 15\mu$K, also follows from the extrapolation of the
CMB power spectrum from higher multipole moments to $l=2$ \cite{bennett96}.

The CMB quadrupole should cause the microwave radiation from clusters of
galaxies to be polarized. Below we present a formalism that
allows one to calculate the quadrupole-induced polarization to be
measured from a cluster with given celestial coordinates for a given
CMB quadrupole. Rather as an example (because of the large uncertainty
present) we find the distribution of the polarization signal over the
sky as implied by the COBE measurement. It will be straightforward to
update our current, rough estimates as new, more accurate measurements
(the first of this kind is likely to be provided by MAP) become
available. Although, the formulae below are directly applicable only for
clusters which belong to the local population, i.e. for measuring the
quadrupole that is observed at the current epoch from Earth, one can
easily extend our approach on the determination of quadrupoles as seen
from highly redshifted clusters.

The COBE quadrupole is defined by 5 parameters, which are essentially
coefficients at spherical second-order harmonics \cite{kogut96}:
$Q_1=(19.0\pm 7.4\pm 8.2)$~$\mu$K, $Q_2=(2.1\pm 2.5\pm 2.7)$~$\mu$K,
$Q_3=(8.9\pm 2.0\pm2.5)$~$\mu$K, $Q_4=(-10.4\pm 8.0\pm4.3)$~$\mu$K,
and $Q_5=(11.7\pm 7.3\pm 10.4)$~$\mu$K. Here, the first quoted errors
are statistical, which are followed by the COBE team's estimates of
the present systematic errors. One can construct from the $Q_i$
parameters a tensor $Q_{\alpha\beta}$ with the following components

\begin{eqnarray}
Q_{xx}=Q_4-\frac{1}{2}Q_1,\,\,\,Q_{yy}=-Q_4-\frac{1}{2}Q_1,\,\,\,Q_{zz}=Q_1,
\nonumber\\
Q_{xy}=Q_{yx}=Q_5,\,\,\,Q_{xz}=Q_{zx}=Q_2,\,\,\,Q_{yz}=Q_{zy}=Q_3.
\label{qi}
\end{eqnarray}
The quadrupole signal, $I_Q$, can be calculated for any direction as
$I_Q=\sum Q_{\alpha\beta} r_{\alpha} r_{\beta}$, where 
the Cartesian coordinates $r_{\alpha}$ are expressed through the Galactic
coordinates $l$ and $b$, i.e. $X=\cos{l}\cos{b}$, $Y=\sin{l}\cos{b}$,
and $Z=\sin{b}$. The $Q_{\alpha\beta}$ tensor possesses the property
$Q_{xx}+Q_{yy}+Q_{zz}=0$.

We can next find the eigenvalues $\lambda_1$, $\lambda_2$,
$\lambda_3$ (let $\lambda_1$ be the minimum and $\lambda_2$ the
maximum among the three $\lambda_i$) and the corresponding
eigenvectors $\vec{e_1}$, $\vec{e_2}$, $\vec{e_3}$ for the
$Q_{\alpha\beta}$ tensor in a usual way: $Q_{\alpha\beta}
\vec{e_i}=\lambda_i \vec{e_i}$. On introducing spherical
coordinates, $\theta$ and $\phi$ (with the polar angle $\theta$
measured from $\vec{e_3}$), we can, using the property
$\lambda_1+\lambda_2+\lambda_3=0$, cast the quadrupole signal in a
simple form:

\begin{eqnarray}
I_Q &=& \lambda_1\sin^2{\theta}\cos^2{\phi}+\lambda_2\sin^2{\theta}\sin^2{\phi}+\lambda_3\cos^2{\theta}
\nonumber\\
    &=& \lambda_x \left(\sin^2{\theta}\cos^2{\phi}-\frac{1}{3}\right)+
\lambda_y \left(\sin^2{\theta}\sin^2{\phi}-\frac{1}{3}\right),
\label{i_q}
\end{eqnarray}
where $\lambda_x=\lambda_1-\lambda_3$ and
$\lambda_y=\lambda_2-\lambda_3$. The RMS amplitude of the quadrupole
signal is given by

\begin{equation}
Q^2_{rms}=\frac{4}{45}(\lambda_x^2-\lambda_x\lambda_y+\lambda_y^2).
\label{q_rms}
\end{equation}
For the COBE specific values (as we use these only for illustration
purposes, we will quote no errors for the numbers to follow), we find
$\lambda_1=-25$~$\mu$K (minimum), $\lambda_2=23$~$\mu$K (maximum),
$\lambda_3=2$~$\mu$K, $\lambda_x=-27$~$\mu$K,
$\lambda_y=22$~$\mu$K. The RMS amplitude $Q_{rms}=12.6$~$\mu$K. Note
that this value was calculated here without taking account of the existing
correlation between the signal and noise in the COBE data, the
inclusion of which would have given a slightly lower value (10.7~$\mu$K),
as explained in \cite{kogut96}. The principal vectors
$\vec{e_1}$, $\vec{e_2}$, $\vec{e_3}$ point to the following positions
in the sky ($l$, $b$): (335\degr, 2\degr), (250\degr, -63\degr),
(64\degr, -27\degr).

We have thus expressed the quadrupole signal as the sum of two
orthogonal components proportional to the Legendre polinomial of
second order, one of which is defined with respect to the minimum
direction of the CMB quadrupole ($\pm\vec{e_1}$) and the other with
respect to its maximum direction ($\pm\vec{e_2}$). We can
therefore take advantage of formulae (\ref{quad_i_tau}) and
(\ref{quad_qu_tau}) of Section 2. Although those equations were
originally thought of in intensity terms, we can carry out our further
analysis in terms of the brightness temperature by introducing
temperature Stokes parameters, ($T$, $Q_T$, $U_T$), bearing in mind
that small changes in temperature parameters, e.g. $\Delta Q_T/T$, are
related to the corresponding changes in intensity parameters,
i.e. $\Delta Q_{\nu}/I_{\nu}$, via the frequency function

\begin{equation}
f(x)=\frac{d\ln{I_{\nu}}}{d\ln{T}}=\frac{x e^x}{e^x-1},
\label{f_k}
\end{equation}
where $x=h\nu/kT$ is a dimensionless frequency. Since the change
in the CMB in the direction of a galaxy cluster is very small, the resulting 
polarization degree will obviously have a frequency dependence defined
by formula (\ref{f_k}), $P^\prime(x)=f(x)(Q_T^{\prime 2}+U_T^{\prime
2})^{1/2}/T$.

After elementary calculations, which are essentially the conversion of
the individual polarization signals that are produced by each of the 
two orthogonal quadrupoles to common axes (see, e.g. Chandrasekhar
1950) and their subsequent summation, we obtain

\begin{eqnarray}
T^\prime(\theta,\phi)=T(\theta,\phi)-\tau\left\{T_1+0.9\left[\lambda_x\left(\sin^2{\theta}\cos^2{\phi}-\frac{1}{3}\right)
\right.\right.
\nonumber\\
\left.\left.
+\lambda_y\left(\sin^2{\theta}\sin^2{\phi}-\frac{1}{3}\right)\right]+\sum_{n=3}^{\infty}T_n\right\},
\label{i_sky}
\end{eqnarray}

\begin{eqnarray}
Q^\prime_T(\theta,\phi)=0.1\tau[\lambda_x(\cos^2{\theta}\cos^2{\phi}-\sin^2{\phi})
\nonumber\\
+\lambda_y(\cos^2{\theta}\sin^2{\phi}-\cos^2{\phi})],
\label{q_sky}
\end{eqnarray}

\begin{eqnarray}
U^\prime_T(\theta,\phi)=0.1\tau(\lambda_x-\lambda_y)\cos{\theta}\sin{2\phi},
\label{u_sky}
\end{eqnarray}

\begin{eqnarray}
P^\prime(\theta,\phi,x)=0.1 f(x) \tau
T_0^{-1}\left\{\left[\lambda_x(1-\sin^2{\theta}\cos^2{\phi})
\right.\right.
\nonumber\\
\left.\left.
+\lambda_y(1-\sin^2{\theta}\sin^2{\phi})\right]^2-4\lambda_x\lambda_y\cos^2{\theta}\right\}^{1/2},
\label{p_sky}
\end{eqnarray}
where $T(\theta,\phi)$ is the initial CMB temperature distribution, in
which the main, isotropic component $T_0=2.73$~K, and $T_1$ and $T_n$ are the
dipole and higher multipole harmonics. In the above, the Stokes
parameters were defined so that if $Q_T>0$ and $U_T=0$, then the
vector of polarization is orthogonal to $\vec{e_3}$.

The polarization degree depends on the cluster position on the
celestial sphere in accordance with formula (\ref{p_sky}). The angular
dependence has two maxima, $P_{max}=0.1 f(x) \tau
T_0^{-1}(\lambda_y-\lambda_x)$, in opposite directions,
$\pm\vec{e_3}$. The maximal polarization signal expressed in absolute
units is $0.1 (\lambda_y-\lambda_x)\tau =4.9\tau $~$\mu$K. 
In the directions of the minima and maxima of the CMB
quadrupole, $\pm\vec{e_1}$ and $\pm\vec{e_2}$, $P=0.1 f(x) \tau
T_0^{-1}\lambda_y$ and $0.1 f(x) \tau T_0^{-1}|\lambda_x|$,
respectively, and these four positions are saddle points in which the
polarization is maximal in the plane $\theta=\pi/2$
and minimal in the perpendicular direction. There are also four
positions in the same $\theta=\pi/2$ plane for which $P=0$:
$\phi=\pm\arctan{(-\lambda_y/\lambda_x)^{1/2}}$ and
$\phi=\pi\pm\arctan{(-\lambda_y/\lambda_x)^{1/2}}$. The geometry of
the effect is sketched in Fig.~1.
\begin{figure}
\plotone{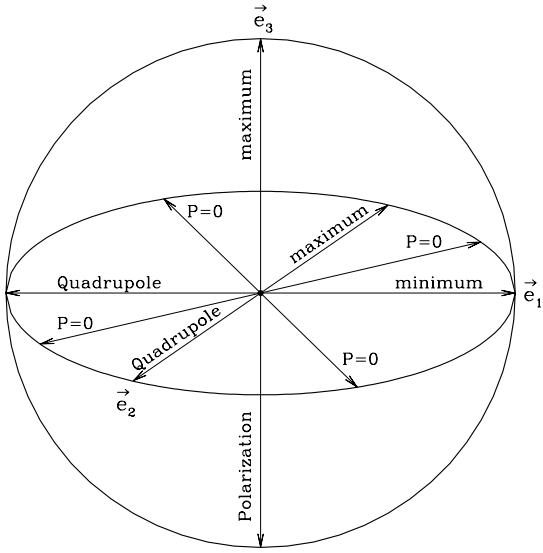}
\caption{The geometry of the polarization effect induced by the
CMB quadrupole. The vectors $\vec{e_1}$, $\vec{e_2}$ and $\vec{e_3}$
define the eigensystem of the CMB quadrupole temperature 
anisotropy. The polarization effect has two broad maxima in the directions
$\pm\vec{e_3}$ orthogonal to the plane which contains the minima
($\pm\vec{e_1}$) and maxima ($\pm\vec{e_2}$) of the quadrupole. In
the same plane there are four directions for which there is no polarization.
} 
\label{fig1}
\end{figure}

The sky-average (RMS) polarization induced by the CMB quadrupole can
be found from expression (\ref{p_sky}) to be

\begin{equation}
P_{rms}(x)=0.1 f(x)\tau
T_0^{-1}\sqrt{\frac{8}{15}(\lambda_x^2-\lambda_x\lambda_y+\lambda_y^2)},
\end{equation}
or, using equation (\ref{q_rms}),

\begin{equation}
P_{rms}(x)=\frac{\sqrt{6}}{10}f(x)\tau\frac{Q_{rms}}{T_0}.
\end{equation}
Note the $\sqrt{6}$ enhancement factor in this expression. For the
COBE data the sky-average polarization is $3.1\tau$~$\mu$K, or $0.6$
of the maximal signal.

Expanding expression (\ref{p_sky}) in the vicinity of the maximum
directions, $\theta=0$, $\pi$, gives

\begin{equation}
\frac{\Delta P}{P_{max}}=\frac{\lambda_x\cos^2{\phi}-\lambda_y\sin^2{\phi}}{\lambda_y-\lambda_x}(\Delta\theta)^2.
\end{equation}
The quadratic dependence on the angular shift implies that the
polarization maxima are very broad. Each of the two roughly elliptical
areas around the maxima within which $P/P_{max}>1/2$ subtends about a
quarter of the sky, as implied by the COBE measurement. Obviously, in
these extended regions one has best chances to detect the
quadrupole-induced polarizion. We note, however, that the measurement
of the effect in the direction of its maxima is not sufficient for
deducing all the 5 parameters that define the global quadrupole, so
measurements in other directions are still necessary. 

In contrast to the maxima, the four directions in which there is no
polarization signal are well resolved, as near them 

\begin{equation}
\frac{P}{P_{max}}=2\frac{\sqrt{-\lambda_x\lambda_y}}{\lambda_y-\lambda_x}\sqrt{(\Delta \phi)^2+(\Delta \theta)^2}.
\end{equation}
The minima are sharp because the Stokes parameters $Q$ and $U$ change
their sign in these points. The area around a minimum within which
$P/P_{max}<1/4$ is a $14^{\circ}$ radius circle for the adopted COBE values.

The polarization caused by the CMB quadrupole is very small, reaching
at the maximum $P_{max}\sim 2\cdot 10^{-6} f(x) \tau$ or, equivalently,
$0.1(\tau/0.02)$~$\mu$K. This fact does not, however, automatically
require the detector sensitivity threshold be 
accordingly low. Instead, one can derive the polarization
corresponding to a given direction by averaging polarization 
signals taken from a large number of clusters located roughly around
this direction (to be more precise, two opposite directions can be
probed at the same time). In such a study, it is, in principle, even not 
necessary to have information on the optical depth of each 
particular cluster. Indeed, the polarization effect depends linearly
on $\tau$, therefore it is the ensemble-average optical depth which is
important. Furthermore, all other possible polarization effects
associated with clusters, which we discuss below, are not additive, in
contrast to the quadrupole effect, i.e. summing up signals from individual
clusters will tend to smooth out their contribution.

\section{OTHER POLARIZATION MECHANISMS IN CLUSTERS OF GALAXIES}

\subsection{Polarization due to cluster transverse motion}

Sunyaev \& Zel'dovich \shortcite{sunyaev_zeldovich80} have shown that
the proper motion of a galaxy cluster relative to the CMB will induce
a polarization in the microwave signal measured from the cluster. They
calculated this effect for the Rayleigh-Jeans region of the CMB
spectrum and indicated (see formula (8) in their 1980 paper) how to
calculate the effect for any desired frequency. We now intend to find
the frequency dependence of the produced polarization in explicit form. 

Let us assume that the initial microwave background is isotropic and
has a black-body spectrum of temperature $T_0$. Now let an electron
move through this background at a speed $v$. In the electron's rest
frame, the radiation spectrum is also black-body, but with a
temperature depending on the viewing direction. The spectral intensity
in this frame is given by

\begin{equation}
I_\nu =C\frac{x^3}{e^{x\gamma(1+\beta\mu)}-1},
\end{equation}
where $C=2(kT_0)^3/(hc)^2$, $\beta=v/c$,
$\gamma=(1-\beta^2)^{-1/2}$, $\mu$ is the cosine of the angle between
the electron's velocity vector and the direction of incidence of a
photon, and all the quantities are measured in the rest frame of the
electron. The angular distribution of the background can be expanded
in Legendre polynomials. Retaining only terms up to the second order in
$\beta^2$, one gets  

\begin{eqnarray}
I_\nu=C\frac{x^3}{e^{x}-1}\left[\left(1+\frac{e^{x}(e^{x}+1)}{6(e^{x}-1)^2}x^2\beta^2\right)-\frac{e^{x}}{e^{x}-1}x\beta\mu
\right.
\nonumber\\
\left.
+\frac{e^{x}(e^{x}+1)}{2(e^{x}-1)^2}x^2\beta^2\left(\mu^2-\frac{1}{3}\right)+...\right].
\end{eqnarray}
Thus the quadrupole component of the initial distribution is

\begin{equation}
\frac{I_2}{I_0}=\frac{e^x(e^x+1)}{2(e^x-1)^2}x^2\beta^2.
\end{equation}
In the limit $x\rightarrow 0$, corresponding to the Rayleigh-Jeans
spectral region, $I_2/I_0=\beta^2$. In the opposite limit,
$x\rightarrow\infty$, one has $I_2/I_0=1/2x^2\beta^2$. The transformation
of the resulting polarization to the CMB frame will produce additional
terms which are of higher than the second order in $\beta$, so we will
ignore them here. Using equations (\ref{quad_i_tau}) and
(\ref{quad_qu_tau}), we then finally obtain the polarization to be
measured towards a cluster moving in the direction $\mu^\prime$ with
respect to the observer,

\begin{equation}
P_\nu^\prime=0.1\frac{e^x(e^x+1)}{2(e^x-1)^2}x^2\beta_t^2\tau.
\label{v2_tau}
\end{equation}
Here, $\beta_t=\beta(1-\mu^{\prime 2})^{1/2}$ is the transverse
component of the peculiar velocity of the cluster. The polarization vector is
perpendicular to the plane formed by the velocity vector and the
observing direction. In the Rayleigh-Jeans frequency region, we readily find
$P_\nu^\prime=0.1\beta_t^2\tau$, exactly the result of Sunyaev \&
Zel'dovich \shortcite{sunyaev_zeldovich80}. In the Wien part of the spectrum
($x\rightarrow\infty$), the degree of polarization of the measured
signal is much higher, a fact mentioned by Zel'dovich \& Sunyaev
\shortcite{sunyaev_zeldovich81}. Fig.~2(a) shows the distribution of
the polarized signal in the picture plane for a spherical cluster.
\begin{figure}
\plotone{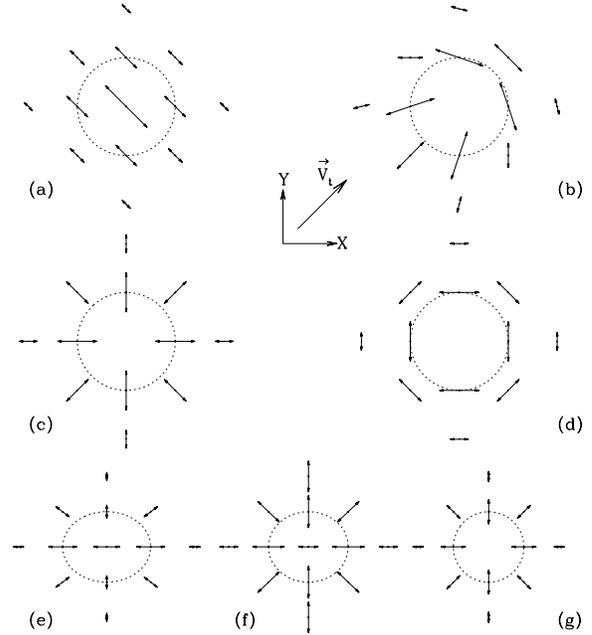}
\caption{Predicted angular distributions of the polarized signal in
the direction of galaxy clusters. In cases (a)--(d) the model gas cloud is
spherical and an isothermal density law with $n=3/2$ is assumed. 
(a) The polarization $\propto\beta_t^2\tau$ caused by the peculiar
motion in the indicated direction (see the insert in the center of the
figure). (b) The finite optical depth polarization
$\propto\beta_t\tau^2$ for the same peculiar motion. (c) The finite
optical depth polarization $\propto(kT_e/m_ec^2)\tau^2$ at frequencies $x<
3.83$. (d) The same but at $x> 3.83$. (e) The gas cloud is a
spheroid with the ratio of the principal axes $b/a=0.8$ and the symmetry 
(longer) axis in the $X$ direction. (f) The same but the symmetry
axis is inclined at $45^\circ$ to the picture plane. (g) The same but
the symmetry axis points along the line of sight. The projected core of the
cluster (as defined by the isothermal law) is shown by the dashed
lines in all panels.} 
\label{fig2}
\end{figure}

It is easy to show that the frequency-integrated polarization 
$\int Q^\prime_\nu\,d\nu/\int I^\prime_\nu\,d\nu=\beta_t^2\tau$,
in agreement with the result of Audit \& Simmons
\shortcite{audit_simmons98}.

\subsection{Polarization effects due to finite optical depth}

Sunyaev \& Zel'dovich \shortcite{sunyaev_zeldovich80} have shown that
there is another polarization mechanism associated with galaxy
clusters which is connected with two consecutive scatterings of a
photon within the intracluster gas, i.e. with finite optical depth of
the gas cloud. Indeed, after first scattering in the intracluster
medium, the CMB acquires an anisotropy due to the thermal and
kinematic effects. The magnitude of this local anisotropic component
is proportional to the product of $\tau$ and $\eta=kT_e/m_ec^2$ for
the thermal effect ($T_e$ being the temperature of the plasma) and
$\beta$ for the kinematic effect. Second scattering within the cluster
will induce polarization signals of the order of $\eta\tau^2$ and
$\beta\tau^2$. In the first approximation, the frequency dependence of
the polarization will be the same as that of the thermal or kinematic
effect. The resulting distribution of the degree and direction of
polarization over the cluster projected on the sky will be 
determined by the distribution of gas density and temperature within
the cluster, on which depends the magnitude of the local quadrupole component. 

Before going further we would like to point out that the thermal
and kinematic effects are not the only possible mechanisms capable of
producing a local anisotropy. The local intensity distribution should also
be distorted by gravitational effects, the most important of which is
the effect of a moving gravitational lense (connected with the cluster
peculiar motion). The resulting polarization was calculated by
Gibilisco \shortcite{gibilisco97}, and for very massive clusters it
can be of comparable magnitude with the effects we discuss.

Sunyaev \& Zel'dovich \shortcite{sunyaev_zeldovich80} estimated the
finite depth effects for a homogeneous spherical gas cloud. We will
first check their results and then consider more realistic models for
the gas distribution in clusters. It turned out that the simple
formulae obtained in Section 2 are not readily applicable to the
problem at hand. We have therefore chosen to follow a more
straightforward approach, which is outlined below.

We introduce a coordinate system in which the $\vec{OZ}$ axis is
pointed from the cluster in our direction and the picture
plane contains the $\vec{OX}$ and $\vec{OY}$ axes. Our task is to calculate the
polarization that would be measured from a given ($X,Y$)
projection point. For a given model we must specify a density law,
$\rho(X,Y,Z)$, where we, for simplicity, by density mean the number density of
electrons. The gas temperature $\eta$ is assumed constant throughout
the cluster in all of our models.

The calculation runs as follows. To the effect from the ($X,Y$)
point contribute all points ($X,Y,Z$) on the line of sight passing
through this projection point. Therefore, one needs to find for each such
point the angular intensity distribution (in which the isotropic
component can be ignored) resulted from the first scattering, $\Delta
I(X,Y,Z,\theta,\phi)$. Here, the $\theta$ and $\phi$ angles define the
viewing direction $\vec{l}$:
$(\sin{\theta}\cos{\phi},\sin{\theta}\sin{\phi},\cos{\theta})$. If we
define the Stokes parameters with respect to the $\vec{OX}$ and $\vec{OY}$ axes
($Q>0$, $U=0$ corresponds to a polarization in the $\vec{OY}$
direction), then the resulting polarization will be

\begin{eqnarray}
\lefteqn{Q^\prime_\nu(X,Y)=\frac{3\sigma_{\rm T}}{16\pi}}
\nonumber\\
\lefteqn{\times \int dZ \rho(X,Y,Z)\int d\vec{\Omega}
\sin^2{\theta}\cos{(2\phi)}\Delta I_\nu(X,Y,Z,\theta,\phi),}
\label{q_tau2}
\end{eqnarray}

\begin{eqnarray}
\lefteqn{U^\prime_\nu(X,Y)=\frac{3\sigma_{\rm T}}{16\pi}}
\nonumber\\
\lefteqn{\times \int dZ \rho(X,Y,Z)\int d\vec{\Omega}
\sin^2{\theta}\sin{(2\phi)}\Delta I_\nu(X,Y,Z,\theta,\phi),}
\label{u_tau2}
\end{eqnarray}
where $\sigma_T$ is the Thomson scattering cross section.

Equations (\ref{q_tau2}) and ({\ref{u_tau2}) take account of the Rayleigh
scattering function and of the fact that the polarization vector of a
photon scattered from the direction $\vec{l}$ towards the observer
makes an angle $\phi$ with the $\vec{OY}$ axis, so the rotation to the
standard axes needs to be implemented (see, e.g. Chandrasekhar 1950).

The intensity $\Delta I_{\nu}$ is equal to the product of the optical depth
$\tau$ as seen from the point $(X,Y,Z)$ in the direction $\vec{l}$ and a
geometry-independent factor which is determined by whether the
kinematic or thermal effect has caused the local anisotropy. For the
thermal effect

\begin{equation}
\Delta I_\nu(X,Y,Z,\theta,\phi)=\tau(X,Y,Z,\theta,\phi)
\eta I_\nu f_{\rm T}(x),
\label{di_t}
\end{equation}
where $I_\nu$ is the CMB spectral intensity, and

\begin{equation}
f_{\rm T}(x)=\frac{x e^x}{e^x-1} \left(x\frac{e^x+1}{e^x-1}-4\right)
\end{equation}
\cite{sunyaev_zeldovich72}. In the case of the kinematic effect,

\begin{equation}
\Delta I_\nu(X,Y,Z,\theta,\phi)=-\tau(X,Y,Z,\theta,\phi)
(\vec{l}\,\vec{\beta}) I_\nu f(x),
\label{di_k}
\end{equation}
where $\vec{\beta}=\beta(\sin{\alpha}\cos{\psi}, \sin{\alpha}\cos{\psi},
\cos{\alpha})$ is the peculiar velocity vector, and the frequency 
dependence for the kinematic effect $f$ is the same as that
for the polarization produced by the CMB quadrupole, given by
formula (\ref{f_k}). Now we proceed to some examples.

\subsubsection{Homogeneous spherical cloud}

\[
\rho(r)= \left\{ \begin{array}{ll}
                \rho_0, & r<1;\\
                     0, & r>1.
                \end{array}
         \right. 
\]

For the $\beta\tau^2$ effect the calculations in equations
(\ref{q_tau2}), (\ref{u_tau2}) can be completed analytically, yielding:

\begin{equation}
Q^\prime_\nu=\frac{1}{20}(X\cos{\psi}-Y\sin{\psi})
\sqrt{1-(X^2+Y^2)}\beta_t\tau_0^2 I_\nu f(x), 
\label{q_k}
\end{equation}

\begin{equation}
U^\prime_\nu=\frac{1}{20}(X\sin{\psi}+Y\cos{\psi})\sqrt{1-(X^2+Y^2)}\beta_t\tau_0^2
I_\nu f(x),
\label{u_k}
\end{equation}
where we have introduced $\tau_0=2\sigma_{\rm T}\rho_0$, the central
optical depth.

The derived result is identical to that of Sunyaev \& Zel'dovich
\shortcite{sunyaev_zeldovich80}. One important property of the effect
described by relations (\ref{q_k}) and (\ref{u_k}) is that its amplitude
($(Q^2+U^2)^{1/2})$ depends only on the projected distance $r$ from 
the cluster center. The azimuthal angle determines only the direction
of polarization; the sign of the effect is different in
the leading and trailing parts of the cluster (with respect to the
direction of motion), as shown in Fig.~2(b).  The maximal
polarization degree is reached at $r_{\rm max}=1/\sqrt{2}$, and is 
equal to $0.025 \beta_t\tau_0^2 f(x)$. In the Rayleigh-Jeans 
region, this is simply $0.025 \beta_t\tau_0^2$. One should note that in
this case, in contrast to the effect of the order of $\beta_t^2\tau$,
there is no net polarization from the cluster.

In the case of the $\eta\tau^2$ effect, one can easily verify that
$U^\prime_\nu(X,Y)=0$. The integration for $Q^\prime_\nu(X,Y)$ in
formula (\ref{q_tau2}) can be implemented numerically. The
polarization direction depends on the frequency of measurement in this
case. At low frequencies, the polarization is radially directed
in the picture plane -- see Fig.~2(c). At $x=3.83$, the thermal
effect changes its sign and at frequencies higher than this critical
value, the polarization vector is orthogonal to the radius-vector, as
shown in Fig.~2(d). The maximal polarization degree of $0.014
\eta\tau_0^2 f_{\rm T}(x)$ is reached at $r=0.85$. In the Rayleigh-Jeans
spectral region, $P=0.028 \eta\tau_0^2$. Again, like for
the $\beta\tau^2$ effect, the angular-integrated polarization is zero.

\subsubsection{Spherical cloud, King's density law}

$\rho(r)=[1+(r/r_c)^2]^{-n}$.

Fig.~3 shows how the polarization degree depends on the angular
distance ($r/r_c$) from the cluster center for two values of $n$:
$1$ and $2$. The maximal polarization is reduced by a
factor of $\sim 2$ compared to the case of a homogeneous sphere for a
given central optical depth $\tau_0$. 
\begin{figure}
\plotone{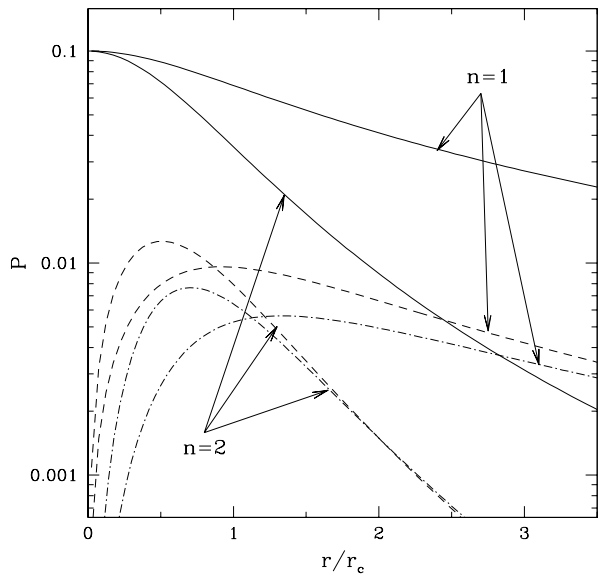}
\caption{Various polarization effects as functions of
projected distance from the cluster center, calculated for the
spherical isothermal density model (for $n=1$ and $2$). Shown is the
factor to be multiplied by: $\beta_t^2\tau_0$ (solid lines; for
this effect as well for the CMB quadrupole effect the angular
dependence simply follows the angular dependence of the line of sight
optical depth), $\beta_t\tau_0^2$ (dashed lines), and
$(kT_e/m_ec^2)\tau_0^2$ (dash-dotted lines), where $\tau_0$ is the
central optical depth.} 
\label{fig3}
\end{figure}

\subsubsection{Ellipsoidal cloud, King's density law}

It is known from observations that the shape of clusters of galaxies
often significantly differs from spherical. The finite optical depth
polarization effects will be modified notably in such situations. Let
us consider an idealized model cluster which has a shape of the
ellipsoid of revolution and whose density distribution is described by a
modified isothermal law:
$\rho(r)=[1+(X/a)^2+(Y^2+Z^2)/b^2]^{-n}$. To make the 
situation more complete, we can rotate this ellipsoid around the $\vec{OY}$
axis. Examples of the resulting polarization pattern are presented in
Fig.~2(e)--(g). The typical polarization degree is, in this case, of
roughly the same magnitude as it was for the spherical cloud model,
but its distribution in the picture plane can be absolutely different. In
particular, generally, there is a polarized radiation in the direction
of the cluster center -- for example in Fig.~2(e) the degree of
ellipsoidality ($b/a$) is only 0.8 but we see a maximum of
polarization near the center, in sharp contrast to Fig.~2(c). Also
important is that the net polarization from the cluster is no longer zero.

It is clear from the above that the angular distribution of
polarization caused by the finite depth effects for a given cluster will
depend strongly on its shape and orientation, as well as on the
distribution of gas within the cluster.

\section{DISCUSSION}
\begin{figure}
\plotone{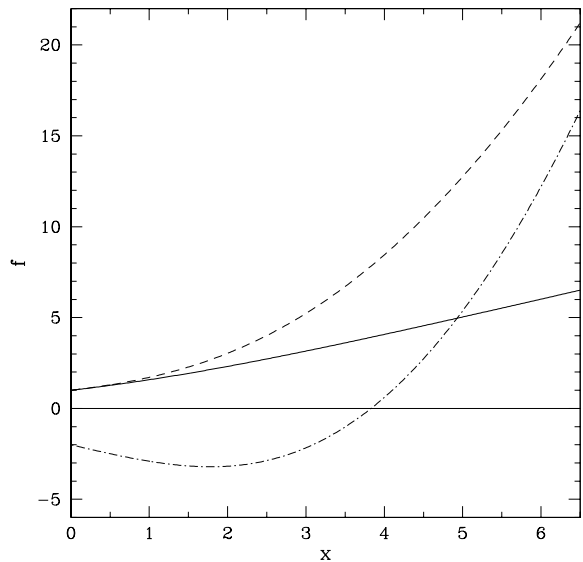}
\caption{Various polarization effects as functions of frequency:
the CMB quadrupole effect and the $\beta_t^2\tau$ effect (solid
line), the $\beta_t\tau^2$ effect (dashed line), and the
$(kT_e/m_ec^2)\tau^2$ effect (dash-dotted line). The latest effect
changes its sign at $x=3.83$.} 
\label{fig4}
\end{figure}

Microwave polarization measurements in directions of galaxy
clusters provide a novel method for determining the quadrupole component
in the CMB angular distribution, which does not require mapping of the
whole sky, as is the case with the conventional method of measuring
the CMB anisotropies. What is more important is that this method
allows one to measure the CMB quadrupole component as would be seen by
observers that are placed on distant clusters, including those at
$z\sim$~0.5--3. This information, which is otherwise unaccessible, has
significant implications for cosmology, which should raise interest of
experimentators to microwave polarization measurements of distant
clusters. Such measurements can at the same time provide another piece
of information valuable for cosmology, namely that on the evolution of
the peculiar velocities of clusters \cite{sunyaev_zeldovich81}. The
most serious concern is, of course, the small magnitude of the
polarization induced by  the CMB quadrupole, which is expected to be
less than $\sim 0.1 (\tau/0.02)$~$\mu$K. However, as we mentioned in
Section 3, one can overcome the detector sensitivity limit by probing
a large number of clusters.

Another problem is the presence of several other polarization effects
connected with clusters of galaxies. These effects, which are
interesting for themselves because they could, in principle, provide us
such important information as on cluster tangential velocities and the
distribution of intracluster gas, can produce a polarization
comparable in size with that due to the CMB quadrupole, and will thus
be a source of significant noise for observations aiming at the
detection of the quadrupole-induced signal. Fortunately, all of 
these effects have a tendency to vanish in the course of averaging
over a large cluster sample. Let us see how significantly the
different polarization mechanisms could contribute to an individual
measurement.

The polarizations due to the three basic effects are of the order
of $0.1 \beta_t^2\tau$, $0.01 \beta_t\tau^2$ and $0.01 \eta\tau^2$ in
the Rayleigh-Jeans spectral region (the coefficients at the last two
effects are model-dependent, but typically not differ much from the
quoted values). The first effect becomes equal to the CMB quadrupole
polarization effect at its maximum for the tangential velocity $\sim
1300$~km/s (as implied by the COBE data). This value is
considerably higher than the average peculiar velocity expected
for clusters of galaxies (see, e.g. Bahcall \& Oh 1996). Moreover, the
sky-average of the polarization signal due to the CMB quadrupole is
$\sim 60$\% of the maximum value, whereas the velocity
effect is proportional to the square of $\beta_t$, and therefore for
typical values, $\beta_t\sim 400$~km/s, the polarization produced by
the CMB quadrupole should be several times stronger in the
Rayleigh-Jeans region of the CMB spectrum. Among the considered
effects, only the $0.01 \eta\tau^2$ polarization might compete with
the quadrupole-induced polarization for rich clusters. 

The above values were calculated for the Rayleigh-Jeans spectral
region, which corresponds to centimeter and millimeter wavelengths. However, 
the polarization degree will strongly depend on frequency if one
performs measurements in the short millimeter and, especially, in the
submillimeter band. The frequency dependencies for the different effects are
shown in Fig.~4. Note that the polarization due to the CMB
quadrupole has the same frequency dependence as the $\beta_t\tau^2$
polarization, both growing linearly ($\sim x$) in the Wien part of the
spectrum. The other two effects rise more steeply, $\sim x^2$, in the Wien
region. The different frequency behaviours provide the opportunity to
separate the signals of different nature by carrying out multi-band
observations. On the other hand, it is clear that there is more hope
to detect the polarization due to the CMB quadrupole in the Rayleigh-Jeans
region, where its contribution is relatively more significant.

The effects being discussed also depend in different ways on the
projected angular distance from the cluster center. The CMB quadrupole
effect and the $\beta_t^2\tau$ effect are proportional to the
optical depth through a given line of sight, and therefore both
lead to a finite polarization of the signal integrated over the
picture plane. The other two effects have a tendency to vanish in the
net signal. This fact significantly enhances our chances to determine
both the CMB quadrupole and the tangential velocities of clusters by
means of polarization measurements -- one just has to accumulate the microwave 
signal from the whole cluster (or at least from an area
which is symmetric around its center). Furthermore, the $\beta_t^2\tau$
polarization will vanish upon averaging over a large set of clusters,
unless the directions of their tangential motions are correlated,
enabling the extraction of the pure quadrupole-induced
signal. Fig.~5 demonstrates the complexity of the overall
problem. It shows how the polarization pattern shaped purely by the 
CMB quadrupole anisotropy will be modified by the other polarization
effects and undergo further changes if one tunes to a different
measurement frequency.
\begin{figure}
\plotone{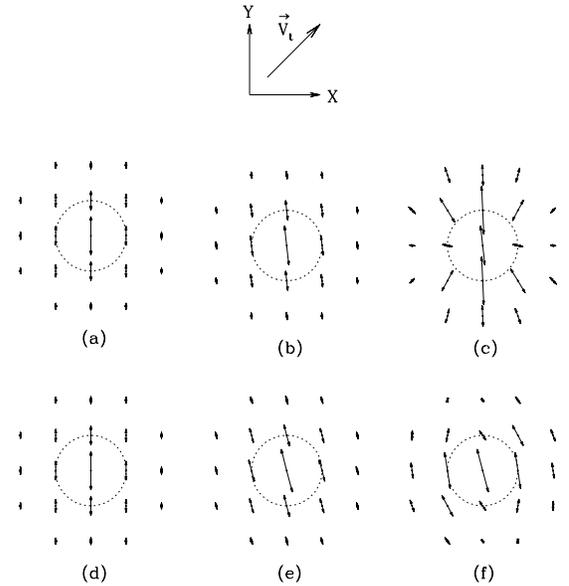}
\caption{Combined effect of the CMB quadrupole and other
mechanisms on the polarization from a cluster which is located in the
direction of the maximum of the quadrupole-induced polarization, has a
transverse velocity of 600~km/s ($\beta_t=0.02$), central optical
depth $\tau=0.01$ and gas temperature 5~keV ($kT_e/m_ec^2=0.01$). The density
distribution is assumed to be $(1+(r/r_c)^2)^{-3/2}$, where $r_c$ is
the core-radius, which defines the size of the circle shown on each
panel. (a) Only the quadrupole-induced polarization. (b) The
$\beta_t^2\tau$ effect has been added. (c) The effect $(kT_e/m_ec^2)\tau^2$ has
been added. Cases (a)--(c) are for the Rayleigh-Jeans spectral
region. (d)--(f) The same as (a)--(c) but for the frequency $x=5$.} 
\label{fig5}
\end{figure}

It is worth noting that other polarization effects, apart from
those discussed in this paper, may appear strong under some
circumstances. For example, the signal from a cluster can be
polarized if there is a strong compact source of unpolarized radio
emission inside the cluster \cite{sunyaev_zeldovich80,sunyaev82}. The
polarization pattern in this case will be similar to that produced by
the $\eta\tau^2$ effect -- see Fig.~2(d). Of course, the frequency
dependence in this case, which should be power-law like, will markedly
distinguish itself. Another way to give rise to polarization is
through gravitational mechanisms \cite{gibilisco97}, which we already
mentioned above. Finally, there is a polarization which is intrinsic
to the CMB. This polarization signal, which together with the
intensity fluctuations has been brought to us largely from the epoch
of decoupling but may also bear imprints from later epochs, is
expected to be very small at sub-arcminute scales which correspond
to the typical dimensions of cluster cores (see, e.g. Seljak \&
Zaldarriaga 1998; White 1998). It should also be noted that the
typical Faraday rotation measures of galaxy clusters are too small to
produce a significant depolarization of the discussed effects for
wavelengths below $\sim 10$~cm, i.e. in the domain of interest to us
\cite{lawler_dennison82,sarazin88}.


\end{document}